\documentclass[a4paper, 12pt, onecolumn]{article}

\title{\bf 
     Family of Commuting Operators for the
     Totally Asymmetric Exclusion Process
}

\author{      O. Golinelli, K. Mallick
\bigskip
\\ \ad        Service de Physique Th\'eorique, Cea Saclay, 91191 Gif, France
}
\date{\normalsize       
                      arXiv:cond-mat/0612351
\\                     v1: 13 December 2006; v2: 06 March 2007
}

  \newcommand  {\ad}{\normalsize\em}      
  \usepackage[final]{graphicx}

  \usepackage{amsfonts} \newcommand{\cc}{\mathbb C} 
  \newcommand{\tr}{\mathrm{tr}}                
  \newcommand{\eqref}[1]{Eq.~(\ref{#1})}       

  \newcommand{\rop}[1]{\mathcal{O}\left(#1\right)} 

  \newcommand{\binomial}[2]{{\scriptstyle #1 \choose \scriptstyle #2 }} 

\begin{document}
\maketitle

\begin{abstract}
  \normalsize
    The  algebraic structure underlying the totally 
      asymmetric exclusion process
     is studied by using the Bethe Ansatz technique.  From  the properties
     of the algebra generated by the local jump
    operators, we explicitly construct the hierarchy 
of  operators (called generalized hamiltonians) 
that commute with the Markov operator.
    The transfer matrix, which is the generating function
    of these operators,  is  shown to represent  a discrete
    Markov process with long-range jumps.  We  give a general
     combinatorial  formula for the 
    {\em connected}  hamiltonians obtained by taking the logarithm
    of the transfer matrix. 
    This formula is proved using a symbolic calculation program for the
    first ten connected operators.

\medskip \noindent 
Keywords: ASEP,  Algebraic  Bethe Ansatz. 

\medskip \noindent 
Pacs numbers: 02.30.Ik, 02.50.-r, 75.10.Pq. 

\end{abstract}

\pagebreak
\section{Introduction}

      The Asymmetric Simple Exclusion Process (ASEP)
   is a driven lattice gas  of 
  particles  that hop   on a lattice  and  interact   through
  hard-core exclusion.  Originally, the ASEP was proposed 
   as a minimal model in one-dimensional
 transport phenomena  with geometric constraints, 
 such as hopping  conductivity, motion of RNA templates and  traffic flow.
  The exclusion process displays a rich phenomenological
 behaviour and  its relative simplicity has  allowed
 to derive many exact  results  in one  dimension. For these reasons,
 the ASEP has become one of the major  models in the
 field of interacting particle systems both in the mathematical
 and the physical literature and   plays the role of a paradigm in  
 non-equilibrium statistical mechanics (for reviews, see e.g.,
 Spohn  1991,  Derrida 1998,  Sch\"utz 2001).

  It has been shown that the evolution
 operator (or Markov matrix) of the exclusion process can be mapped
 into a non-hermitian Heisenberg spin chain of the XXZ type
  (Gwa and Spohn, 1992; Essler and Rittenberg 1996). This 
  mapping allows the  use  of the techniques of integrable
 systems  such as the coordinate Bethe Ansatz (for a review see, e.g.,
 Golinelli and Mallick, 2006b). 
  Spectral information about the
  evolution operator (Dhar 1987; Gwa and Spohn 1992; Sch\"utz 1993;
   Kim 1995;  Golinelli and Mallick 2005)  and  large deviation
  functions (Derrida and Lebowitz  1998) can be derived with the help
 of coordinate  Bethe Ansatz. 
  Besides, using 
  the  more elaborate  algebraic  Bethe Ansatz  technique, 
   the eigenstates of the  Markov matrix can be 
   represented as  Matrix Product states over 
  finite dimensional  quadratic  algebra (Golinelli and Mallick, 2006a).
 The  algebraic  Bethe Ansatz  also plays a fundamental role
  in the derivation of  the Bethe equations for  ASEP with open boundaries 
   (de Gier and Essler, 2005, 2006).

  The aim of the present work is to explore  the algebraic
 properties of the totally asymmetric  exclusion process  
(TASEP) that stem  from
 the algebra generated by the local jump operators that 
 build the Markov  matrix. The  algebraic  Bethe Ansatz technique
   allows   to construct a hierarchy of generalized hamiltonians that 
   contain the  Markov  matrix and commute with each other.
 The generating operator for this family, called 
  the transfer matrix,   defines therefore  a   commuting family of operators
 that can be simultaneously diagonalized.
  We derive,   using the local jump operators algebra, 
  explicit  formulae
 for the transfer matrix and the generalized hamiltonians 
  and characterize their action on the configuration space. 
  These  generalized hamiltonians are non-local because they act
  on non-connected  bonds of the lattice.
  However,  connected operators  
  are generated by taking
  the logarithm of the  transfer matrix.  We study these
   connected operators  and give an explicit formula for them.  

 The outline of this work is as follows~: in Section 2,
 we describe the basic algebraic properties of the 
 totally asymmetric exclusion process
 and define the associated algebra. In Section 3,
 we  give  explicit formulae for the transfer matrix and for the
 generalized hamiltonians that commute with the Markov matrix $M$.
 In particular, we show that the transfer matrix can be interpreted
 as a  discrete time Markov process and we describe the  non-local actions 
 of  the   generalized  hamiltonians. In Section 4,
 we study the connected operators obtained by taking the logarithm
 of the  transfer matrix and propose a  conjectured general formula
 for these  local  operators. The actions of  these operators
 are described  explicitly.   Some mathematical  proofs are given in
 the appendices. 

\section{Algebraic properties of the TASEP}

\subsection{Definition of the model}

\begin{figure}[th]
  \center{\includegraphics[height=5.0cm]{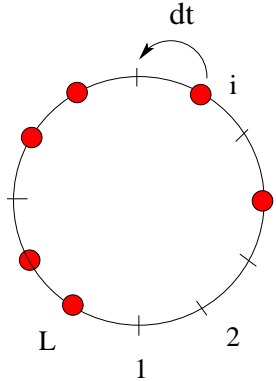}}
   \caption{The totally asymmetric exclusion process on a 
 ring. Sites are labelled from 1 to $L$;  a particle jumps
 with probability $dt$ to the neighbouring forward site if this
 site is vacant.}
  \label{fig:periodicASEP}
\end{figure}

The simple exclusion process is  a continuous-time Markov
process ({\it i.e.},  without memory effects)  in which
 indistinguishable particles  hop  from one site to another
 on a  discrete lattice and obey  the  {\em exclusion
rule}  which  forbids to have more than one  particle  per   site.
 In this work, we  shall study   the case of particles hopping on 
 a  periodic 1-d ring  (see figure~\ref{fig:periodicASEP})
  with $L$ sites labelled $i=1,\dots,L$ 
 (sites $i$ and $i+L$ are identical due to  periodic boundary
conditions).
The particles move   according to the following dynamics: during the time
interval $[t, t+dt]$,  a particle  on a site $i$   jumps with
probability $dt$ to the neighboring site $i+1$,  if this site  is empty.
 This model  is called `totally asymmetric'
 because the particles can jump  only in  one direction.  
 The exclusion rule forbids    particles to  overtake
 each other  and their ordering remains unchanged. Moreover, 
 as  the system is  closed, the number of particles is constant.

The state of a  site $i$ is  encoded in a boolean  variable $\tau_i$,
 where  $\tau_i =1$  if $i$ is occupied and  $\tau_i =0$ otherwise.
   The  two-dimensional state  space of the site $i$
is noted $V_i$ (we have $V_i = \cc^2$)  and its  basis is given by 
 $(|1_i\rangle, \ |0_i\rangle)$.
 A configuration $C$ of the system of $L$ sites is written as 
\begin{equation}
 C = |\tau_1,  \tau_2  \dots  \tau_L \rangle \, .
 \label{conf}
\end{equation}
 The state space  $\mathcal{H}_L$ of the ring is therefore a  $2^L$
 dimensional vector space given by
\begin{equation}
  \mathcal{H}_L = V_1 \otimes V_2 \otimes \dots \otimes V_L
  \, . 
\end{equation} 
 Due to  the conservation of the number $n$  of particles,
 $\mathcal{H}_L$  splits into invariant spaces 
 ${\mathcal H}_L^{(n)}$  of dimension $L! / [ n! (L-n)!]$,
 characterized by $\sum_{i=1}^n\tau_i = n$.

The probability distribution of the system  at time $t$  can be
 represented as a vector  $\psi(t) \in \mathcal{H}_L$, where
 the component $\psi_C(t)$ is the  probability of being in 
  the configuration   $C$ at time $t$.   The vector  
 $\psi(t)$ evolves according to
the {\em master equation}
\begin{equation}
  \frac{d\psi(t)}{dt} = M \psi(t)  \, , 
  \label{markov}
\end{equation}
where $M$ is  a  $2^L \times 2^L$ Markov matrix acting on
$\mathcal{H}_L$.  For $C \ne C'$,   $M(C',C)$ is the transition
rate from configuration $C$ to configuration $C'$: it is equal to 1 if
$C'$ is obtained from $C$ by an allowed jump of a  particle, and 0
otherwise. The diagonal elements are negative and $-M(C,C)$ 
is  the exit
rate from $C$, {\it i.e.},  the number of allowed jumps from $C$.
The  sums over columns  of $M$ vanish,   
$\sum_{C'} M(C',C)= 0 $. This property
 ensures  probability conservation~: 
$\sum_C \psi_C(t) = \sum_C \psi_C(0)= 1$.

In the case of  the  TASEP on  a periodic ring,
 sums over rows of $M$ also vanish. This implies that the
stationary probability, obtained for $t \to \infty$,  is uniform
 over each  subspace ${\mathcal H}_L^{(n)}$.

\subsection{The algebra of jump matrices}
\label{section:mi}

The Markov matrix can be written as 
\begin{equation}
   M = \sum_{i=1}^L M_i \, , 
   \label{msm}
\end{equation}
where  the local jump operator
 $M_i$ represents the contribution  to the dynamics
 of jumps from the site $i$ to
$i+1$.  Thus, the action of the $2^L \times 2^L$ operator
$M_i$   affects only  the  sites $i$ and $i+1$ and is non-zero  only if
$\tau_i = 1$ and $\tau_{i+1} = 0$~:
\begin{eqnarray}
   M_i |\tau_1  \dots 1, 0  \dots \tau_L\rangle &=&
      |\tau_1  \dots 0, 1  \dots \tau_L\rangle 
    - |\tau_1  \dots 1, 0  \dots \tau_L\rangle \ ,
\\
   M_i|\tau_1  \dots \tau_i,  \tau_{i+1}  \dots \tau_L\rangle &=& 0,
    \ \ \ \ \mbox{if} \ \tau_i = 0 \ \mbox{or} \ \tau_{i+1} = 1 \, .
\end{eqnarray}
 The operator $M_L$ corresponds to jumps from site $L$ to 1 (the
  site $L+1$ is identical  to the site 1 because of the 
 periodic boundary  conditions).  

 Using this definition of the local jump operators, it can be 
 verified that the  $M_i$ satisfy  the following  relations~:
\begin{eqnarray}
   M_i^2 &=& - M_i ,
   \label{mi2}
\\    
   M_i \ M_{i+1} \ M_i = M_{i+1} \ M_i \ M_{i+1} &=& 0 ,
   \label{mmm=0}
\\
   \left[M_i,M_j\right] &=& 0 \ \ \ \mathrm{if} \ |i-j| > 1 .
   \label{mimj}
\end{eqnarray}

As we consider a {\em periodic} system, we use the convention $M_{L+1}
\equiv M_1$ in the above relations.  We emphasize that  $[M_L,M_1] \ne 0$ and
$M_1 M_L M_1 = M_L M_1 M_L = 0$.

 The algebra generated 
 by the   $M_i$ operators will be called here the {\it TASEP algebra}.  
  We remark that the $M_i$  operators  can be obtained as a limit of the
Temperley-Lieb algebra  generators.
 We shall call, by definition, a   {\it word},  
 any  product of the  $M_i$'s;  
 any element of the algebra can be written as  a linear
combination of words. The  {\it length} of a word is the minimal
   number of  operators $M_i$ required to write it. 

Each word  acts on the configuration space $\mathcal{H}_L$
 and can be described as  a series  of jumps.  For example, the word
$M_1 M_2$ describes a jump of a particle from site 2 to 3, followed  by
a jump of another particle from site 1 to the site 2; the action
 of $M_1 M_2$ on a configuration   vanishes
 unless $\tau_1 = 1, \,   \tau_2 =1, \,   \tau_3 =0$ and  we have 
\begin{equation}
   M_1 M_2 |1, 1, 0, \tau_4  \dots \tau_L \rangle  =
  | 0, 1, 1,\tau_4 \dots \tau_L \rangle  -
 |1, 0, 1,\tau_4 \dots \tau_L \rangle \, . 
\end{equation}
  Similarly, the  word $M_2 M_1$ represents  a jump of a
particle from site 1 to 2  followed by  a jump of the same particle from
2 to 3~:
\begin{equation}
  M_2  M_1  |1, 0, 0, \tau_4  \dots \tau_L \rangle  =
  | 0, 0, 1,\tau_4 \dots \tau_L \rangle 
   - |0, 1, 0, \tau_4 \dots \tau_L \rangle \, . 
\end{equation}
 Clearly, $M_1$ and $M_2$ do not commute because the jumps on two
 adjacent sites are not independent.

\subsection{Ring-ordered  product of jump matrices}
\label{section:op}

 We  define  here  the  ring-ordered  product of jump  matrices
  which will be used   in the  following sections.

  The  ring-ordered product $\rop{}$  acts on 
words   of the type   
\begin{equation}
 W = M_{i_1} M_{i_2} \dots M_{i_k} \, \hbox{   with  } \, 
  1 \le i_1 < i_2 < \dots < i_k \le L \, ,
 \end{equation}
   by   changing the  positions of matrices that appear in  $W$ 
   according to the following rules~:

     (i)  If $i_1 > 1$ or $i_k < L$, we define  $\rop{W} \ = W$. The  word
 $W$  is  well-ordered.

     (ii) If $i_1=1$ and $i_k=L$,
  we first   write  $W$ as a product of two blocks, $W = AB$,
  such that  $B = M_b
M_{b+1} \dots M_L$ is the maximal block of matrices with consecutive indices 
 that  contains  $M_L$,   and $A = M_1 M_{i_2} \dots M_{i_a}$, 
 with    $i_a < b-1$,  contains the  remaining terms. 
   We  then define
\begin{equation}
   \rop{W} =  \rop{AB} = BA 
        = M_b M_{b+1} \dots M_L M_1 M_{i_2} \dots M_{i_a} \, .
   \label{abba}
\end{equation}

  (iii) The previous  definition  makes sense  only for $k<L$.  Indeed, when 
  $k=L$, we have $W =M_1 M_2 \dots M_L$  and it is not possible 
  to split  $W$ in two different  blocks $A$ and $B$.
   For  this  special case,  we  define
\begin{equation}
  \rop{M_1 M_2 \dots M_L} \ = 
   | 1, 1, \dots, 1 \rangle \langle 1, 1, \dots, 1| \, , 
  \label{m1ml}
\end{equation}
 which  is  the projector on the   `full'   configuration with  all sites 
occupied.
 
 The ring-ordering  $\rop{}$  is extended by  linearity to  the vector space
 spanned by words of the type described above. 

   Let us give some examples.
  For $k=0$ or 1, the ring-ordered product acts trivially: $\rop{1} =
1$ and $\rop{M_i} = M_i$.  For $k=2$ , we have 
$ \rop{M_i M_j} =  M_i M_j $ when $1 \le i < j \le L$; however,
 for the special case 
 when  $i=1$ and $j=L$,  $\rop{M_1 M_L}  =  M_L M_1$.

 The ring-ordered  product  embodies  the periodic boundary conditions.
 On a ring,  the 
 natural order between integers is not  valid. Indeed,
  $M_L$ and $M_1$  act as  neighbouring bonds and 
 site $L$  should be viewed as being
  'behind' site 1, just as site 1 is  behind site 2. The  ring-order
 product  restores  the correct  order on a ring and allows 
  to construct operators  that are  translation invariant.
 For example, for $L=3$, the operator
 $ U = M_1M_2 + M_2M_3 + M_1M_3$   is not well-ordered and
does  not commute with  translations.  But, 
 $\rop{U} =  M_1M_2 + M_2M_3 + M_3  M_1  $ 
 is  well-ordered  and  does  commute with  translations.
 Finally,  we remark that when a ring-ordered product acts
 on a configuration, each particle advances  by at most 
 one lattice unit~: indeed, because  terms such as 
 $M_{i+1} M_i$ do not appear  in a ring-ordered product, no particle
 can perform multiple jumps.

\section{Transfer matrix and  generalized hamiltonians}

The algebraic Bethe Ansatz is a method for diagonalizing the hamiltonian
of integrable models (for a review, see, e.g., Korepin et al. 1993; 
for a pedagogical introduction, see, e.g., Nepomechie 1999).
This technique 
   can be applied to  the Markov matrix $M$ of
  the TASEP (Golinelli and Mallick, 2006b).  The key step is to construct 
  a family of  transfer matrices,   $t(\lambda)$,
 which  act  on the configuration space $\mathcal{H}_L$.  
 For  any value  $\lambda$ and $\nu$ of the spectral parameter, 
 we   have
 \begin{equation}
  [ t(\lambda), t(\nu) ] = 0 \, .
  \label{tt}
\end{equation}
   Thus,  the operators  $t(\lambda)$  form a one-parameter
  family of commuting operators
  which depend  on  a real number $\lambda$, called the  spectral parameter.
 This family contains the Markov matrix $M$ as will be shown below.
  Therefore, all the  $t(\lambda)$'s   share with $M$  a
{\em common} eigenvector basis   independent of $\lambda$.  For the
TASEP, these eigenvectors are determinants of matrices involving
the roots of the  Bethe equations  and the 
  corresponding  eigenvalues are functions  of $\lambda$ 
 (see e.g., Golinelli and Mallick 2006b for an  explicit  formula). 

 The transfer matrix  $t(\lambda)$ is a polynomial
  in  $\lambda$ of degree $L$~: we 
can thus define $H_1, H_2, \dots H_L$  as follows
\begin{equation}
     t(\lambda) = t(0) \ \left( 1 + \sum_{k=1}^L \lambda^k H_k \right) \ .
   \label{thk}
\end{equation}
 The   $H_k$  operators  are $2^L \times 2^L$ matrices
 acting   on the  configuration
space $\mathcal{H}_L$. The   $H_k$'s  will be called  `generalized
hamiltonians' by analogy with quantum spin systems  (Arnaudon
et al. 2005). As the $H_k$'s  are derivatives of $t(\lambda)$,
they also commute with each other~:
\begin{equation} 
 [H_k, t(\lambda)] = 0 , 
 \ \ \ \ 
 [H_j, H_k] = 0 
 \label{hthh}
\end{equation}
for all $j$, $k$ and $\lambda$.  More generally, any  operator  generated
from  $t(\lambda)$, or equivalently from the generalized
    hamiltonians, belongs to the  same commuting family.

  The above considerations are familiar in the framework of algebraic
  Bethe Ansatz.  In Appendix~\ref{appendix:aba},  we explain
   how  the transfer matrix  can be constructed
  using this method.

\subsection{Expressions of the $H_k$'s}

 In this section  we  describe  our results  which are
 specific to the TASEP and give explicit formulae  for 
 the generalized hamiltonians $H_k$. The calculations 
 leading to these  expressions 
 are carried out in detail  in Appendix~\ref{app:calculHk}.
 
 The  $t(0)$ operator  appearing  in equation~(\ref{thk})
  is the {\em translation operator} on the ring and is defined as 
\begin{equation}
   t(0) | \tau_1, \tau_2, \dots, \tau_L \rangle =
   | \tau_2, \dots, \tau_L, \tau_1 \rangle \ .
   \label{pt0}
\end{equation}

 The operator $H_1$  given by
\begin{equation}
   H_1 = t'(0) / t(0) = \sum_{i=1}^L M_i = M \, , 
\end{equation}
  is precisely the Markov matrix $M$  which 
  thus belongs to the commuting family generated by $t(\lambda)$. 

  All the   $H_k$'s  can  be explicitly 
calculated.  By using the ``ring-ordered product'' defined in
section~\ref{section:op}, we find in Appendix~\ref{app:calculHk} that 
 for $1 \le k \le L$,
\begin{equation}
   H_k = \sum_{1 \le i_1 < i_2 < \dots < i_k \le L} 
          \rop{M_{i_1} M_{i_2} \dots  M_{i_k}}
   \ .
   \label{hk}
\end{equation}
 In particular, we have 
  \begin{equation}
    H_2 = \sum_{1 \le i < j \le L} \ \rop{M_i M_j} \ ,
    \label{h2}
  \end{equation}
and  according to equation~(\ref{m1ml}),  
\begin{equation}
  H_L \ = 
   | 1, 1, \dots, 1 \rangle \langle 1, 1, \dots, 1| \ .
  \label{hl}
\end{equation}
   For $k<L$, all the  terms  in  $H_k$ are   products of $k$ jump matrices
 which, because of ring-ordering,  correspond to 
 $k$ {\em different} particles jumping simultaneously one step forward. 
   Thus, $H_k$ has  a non-vanishing  action only on
     configurations with at least $k$ particles.

In the case with  $L=4$, for example,  the generalized hamiltonians are
given by
\begin{eqnarray}
    H_1 &=& M_1 + M_2 + M_3 + M_4 = M \ ,
\\
   H_2 &=& M_1 M_2 + M_2 M_3 + M_3 M_4 + M_4 M_1 + M_1 M_3 + M_2 M_4 \ ,
\\
   H_3 &=& M_1 M_2 M_3 + M_2 M_3 M_4 + M_3 M_4 M_1 + M_4 M_1 M_2 \ .
   \label{l4}
\end{eqnarray}

Using Eqs.~(\ref{thk}) and  (\ref{hk}),  we conclude that 
 the generating function of the
$H_k$  is given by 
\begin{equation}
   t_g(\lambda) =  \frac{ t(\lambda)}{t(0)}=  1 + \sum_{k=1}^L \lambda^k H_k 
    =  \rop{\prod_{i=1}^L (1 + \lambda M_i)} \, . 
   \label{tg}
\end{equation}
 Although
 the  operator  $H_1$ is the Markov matrix $M$ of the TASEP, 
  we emphasize that
   when $k \ge 2$, 
 $H_k$ cannot be interpreted  as a  Markov matrix because 
 it  contains  negative  non-diagonal matrix elements.
 However,   we  shall now  prove   that  
 the matrix $t(\lambda)$  is  the Markov  matrix
 of    a discrete  time  process when  $0 \le \lambda \le 1$.

\subsection{Action of the transfer matrix on a given configuration}
\label{sec:action}

  We describe now the  action  of 
$t(\lambda)$  on a given configuration $|\tau_1, \tau_2,
\dots, \tau_L \rangle$. Using equation~(\ref{tg}), we observe that
\begin{equation}
  t_g(\lambda)|\tau_1, \tau_2,\dots, \tau_L \rangle
=   \rop{\prod_{\stackrel{i=1}{\tau_i=1}}^L  (1 + \lambda M_i) }
 |\tau_1, \tau_2,\dots, \tau_L \rangle \, , 
   \label{tg2}
\end{equation}
where the product runs only over occupied sites.
 This expression shows that the action
 of  $t_g(\lambda)$ is  factorized block by block. 
 We  consider first  the simple block   $ |0 1^p  \rangle$  
  (the  notation $| 0  1^p\rangle$  
  means  that  a  hole is  followed by  $p$ particles).
 We have 
 \begin{displaymath}
 t(\lambda) |0 1^p  \rangle =  t_g(\lambda)  | 1^p 0  \rangle = 
(1 + \lambda M_1) \ldots (1 + \lambda M_p) | 1^p 0  \rangle
= \sum_{k=0}^{p} f_{k,p}  | 1^k 0 1^{p-k} \rangle 
 \, ,  
 \end{displaymath}
 \begin{equation}
   \hbox{ with }  f_{0,p} = \lambda^p  \,  \hbox{ and } \, 
  f_{k,p} = (1-\lambda)  \lambda^{p-k} \hbox{ for } 1 \le k \le p \, .
  \label{t10}
 \end{equation}
 More generally, for a configuration  $C$ of the form
 $| 0^{q_1} 1^{p_1} 0^{q_2}  1^{p_2} \dots 0^{q_s} 1^{p_s} \rangle $ 
 with $p_i,q_i \ge 1$,  we obtain
 \begin{eqnarray}
  \lefteqn{ t(\lambda)C   = t_g(\lambda)
  | 0^{q_1-1} 1^{p_1}  \dots  0^{q_s} 1^{p_s} 0\rangle  = }
  \nonumber \\
&&  | 0^{q_1-1}\rangle  \otimes
 \left( \sum_{k_1=0}^{p_1} f_{k_1,p_1}  | 1^{k_1} 0 1^{p_1-k_1} \rangle\right)
    \otimes | 0^{q_2-1}\rangle  \otimes \ldots   
\nonumber \\ &&
  \otimes  
  \left( \sum_{k_s=0}^{p_s} f_{k_s,p_s}  | 1^{k_s} 0 1^{p_s-k_s} \rangle\right)
 \, .  
\label{eq:actionT}
\end{eqnarray}

Except for the full configuration $|1^L\rangle$ for which $ t(\lambda)
|1^L\rangle = t_g(\lambda) |1^L\rangle = (1+\lambda^L) \ |1^L\rangle $
and the void configuration $|0^L\rangle$ for which $t(\lambda) \
|0^L\rangle = t_g(\lambda) \ |0^L\rangle = |0^L\rangle$, any
configuration $C$ has at least one particle and one hole.  By using the
translation operator $t(0)$ that  commutes with  $t(\lambda)$,
 it is possible to bring  $C$ to a  form  to 
  which  equation~(\ref{eq:actionT})  can be applied. 

We notice that $t(1)$ is the identity operator.  Consequently $t_g(1)$
is the forward translation operator, $t_g(1) = t(0)^{-1}$.

We illustrate these  results  with an
example of  3 particles on a ring of 5 sites: 
\begin{eqnarray}
  \nonumber
  \lefteqn{
     t(\lambda)   \ |10101\rangle =
     t_g(\lambda) \ |01011\rangle =
  } \\ & &
     (1-\lambda)^2           \ |01011\rangle +
     \lambda   (1-\lambda)   \ |00111\rangle 
     + \lambda   (1-\lambda)^2 \ |11010\rangle   
  \nonumber \\ & & 
     +  \lambda^2 (1-\lambda)   \ |10110\rangle 
     + \lambda^2 (1-\lambda)   \ |11001\rangle +
     \lambda^3               \ |10101\rangle \ . \,\,\,\, 
\end{eqnarray}

By considering the action of the operators $t(\lambda)$ and
$t_g(\lambda)$, we remark that for $ 0 \le \lambda \le 1$, $ f_{k,p} \ge
0$ and that $\sum_k f_{k,p} = 1$. The quantities $ f_{k,p}$ can thus
interpreted as probabilities.  The operators $t(\lambda)$ and
$t_g(\lambda)$ are then Markov matrices of discrete time exclusion
processes with parallel dynamics, in which different holes can jump
simultaneously through clusters of particles.

With $t(\lambda)$, a hole located on the left of a cluster of $p$
particles can jump a distance $k$ in the forward direction, $1\le k \le
p $, with probability $\lambda^{p-k}(1-\lambda)$.  The probability that
this hole does not jump at all is $\lambda^p$.

With $t_g(\lambda)$, a hole located on the right of a cluster of $p$
particles can jump a distance $k$ in the backward direction, with
probability $\lambda^{k}(1-\lambda)$ for $1\le k < p $, and with
probability $\lambda^p$ for $k=p$.  The probability that this hole does
not jump at all is $1-\lambda$.

The $t_g(\lambda)$ Markov process is equivalent to a 3-D anisotropic
percolation model and a 2-D five-vertex model (Rajesh and Dhar 1998).
It is also an adaptation on a periodic lattice of the ASEP with a
backward-ordered sequential update (Rajewsky et al. 1996, Brankov et
al. 2004), and equivalently of an asymmetric fragmentation process
(R\'akos and Sch\"utz 2005).  Consequently Markov matrices of these
models on a periodic lattice form a commutating family.

\subsection{Invariance properties of  the transfer matrix}

We  describe here   the symmetries  of  the transfer matrix
  $t(\lambda)$ and
of the  operators $H_k$.  
 Translation invariance   is obvious 
 because   $t(0)$  is the translation operator and  commutes
 with $t(\lambda)$ and $H_k$.
 From  Eqs.~(\ref{hk}, \ref{tg}),   
  we observe    that   $t(\lambda)$ and $H_k$   conserve
  the number $n$  of particles because each jump matrix $M_i$  does so.  
  For a given value of $n$, $t(\lambda)$  is a polynomial of  degree $n$.

  The Markov matrix $M$ is symmetric under  
lattice  reflection  $R$ 
(obtained by 
 exchanging sites  $i$ and   $L-i+1$) followed by  particle-hole
conjugation  $C$ (Golinelli and Mallick 2004). This 
$CR$ symmetry acts on  a configuration  as follows
\begin{equation}
  CR | \tau_1, \tau_2, \dots, \tau_L \rangle =
   | 1-\tau_L, \dots, 1-\tau_2, 1-\tau_1 \rangle \, . 
\end{equation}
 The $CR$ symmetry
  does not
commute with the translation operator $t(0)$ because $CR \ t(0) =
t(0)^{-1} \ CR$. The following property
\begin{equation}
  CR \ M_i \ CR = M_{L-i} \, , 
  \label{cr-mi-cr}
\end{equation}
 implies that  $CR$ is  a symmetry of the Markov matrix,  {\it i.e.},
  $(CR)M(CR) = M$. However, 
$CR$ is {\em not} a symmetry of $H_k$ for $k \ge 2$ because the
orientation of matrices along the ring is inverted by \eqref{cr-mi-cr}.
More precisely,  $H_k$ and $t(\lambda)$ are transformed as follows
\begin{equation}
  \tilde{H}_k = CR \ H_k \ CR \ , 
  \ \ \tilde{t}(\lambda) = CR \ t(\lambda) \ CR \ ,
\end{equation}
where $\tilde{H}_k$ and $\tilde{t}(\lambda)$ are given by formulae
similar to Eqs.~(\ref{hk}, \ref{tg}) but with an {\em anti-}ring-ordered
product $\tilde{\mathcal{O}}$ instead of the  ring-ordered product
$\mathcal{O}$.    With the   $\tilde{H}_k$ operator, 
 $k$ different  holes  jump simultaneously one step
  backward.  For  $k \ge 2$, one can verify that 
  the action  of  $\tilde{H}_k$
and  of $H_k$  on a given configuration   are different.

The $CR$ symmetry allows us  to construct two different families
  $t(\lambda)$ and $\tilde{t}(\lambda)$ of commuting operators, 
 i.e.,   $[t(\lambda), t(\nu)] = 0$ and
$[\tilde{t}(\lambda), \tilde{t}(\nu)] = 0$.  Both families
 contain the  Markov matrix $M = H_1 = \tilde{H}_1$.
 However,  $t(\lambda)$ and
$\tilde{t}(\nu)$ do not commute with each other  for generic
 values of $\lambda$ and $\nu$. 

\section{Connected Operators}

  In the previous section,  we have 
  defined  a set of commuting operators, the
 generalized hamiltonians $H_k$,  that  act  on $k$ different
particles.   However,  these actions are generally not local because they
 involve  particles with arbitrary distances  between them.
Moreover, as can be seen from  \eqref{hk},  
   the number of terms in $H_k$ 
 for a large system of size $L$ grows as $L^k/k!$.  In
  statistical physics,    quantities  that are local and extensive
 are preferred.  Such   ``connected''
(or local) operators are  usually built   from   the logarithm of the
generating function (L\"uscher, 1976). 
 Therefore, for   $k\ge 1$, we  define the {\it
connected hamiltonians} $F_k$   
 as follows~: 
\begin{equation}
  \ln t_g(\lambda) = \sum_{k=1}^{\infty} \frac{\lambda^k}{k} F_k \ .
  \label{lntg}
\end{equation}

The $F_k$'s can be expressed from the $H_k$'s using
definition~(\ref{tg}).  Because the $H_k$'s are {\em commuting}
matrices, the $F_k$'s are also a set of commuting operators and moreover
they commute with $t(\lambda)$ and with all the $H_k$'s, i.e. $[F_k,
H_j] = 0$.  Consequently the $F_k$'s can be calculated with the usual
moments-cumulants transformation,
\begin{equation}
  F_k = k H_k - \sum_{i=1}^{k-1} F_i \ H_{k-i}
  \label{fkhk}
\end{equation}
which is obtained from the derivative of $\ln t_g(\lambda)$.

 We  now  show  that  $\ln
t_g(\lambda)$  and   the $F_k$'s  are linear combinations
 of   connected words,  {\it i.e.},  words which
cannot be factorized in two (or more) commuting words. 
Consider a  word  $W$
 of $\ln t_g(\lambda)$  made of jump matrices $M_i$ with
$i \in \mathcal{I}\subset \{1,2,\dots,L\}$. This word  must
 also appear in   $\ln t_{\mathcal{I}}(\lambda)$ with 
\begin{equation}
   t_{\mathcal{I}}(\lambda) 
       = \rop{\prod_{i \in \mathcal{I}} ( 1 + \lambda M_i)} \ .
\label{logtI}
\end{equation}
Assume that the set of indices  $\mathcal{I}$
  can be split  into  two disjoint  subsets, $\mathcal{I}_1$
and $\mathcal{I}_2$, such that $[M_a, M_b] = 0 $ for all $a \in
\mathcal{I}_1$ and all $b \in\mathcal{I} _2$.
Then the ring-ordered product in equation~(\ref{logtI})
  can be factorized  in two non-connected products and we have 
\begin{equation}
  \ln t_{\mathcal{I}}(\lambda) 
       = \ln t_{\mathcal{I}_1}(\lambda) + \ln t_{\mathcal{I}_2}(\lambda)
       \ .
  \label{i1i2}
\end{equation}
Therefore $W$ must be made of  jump matrices $M_i$ with 
 indices $i$  all belonging either to $\mathcal{I}_1$ or to  $\mathcal{I}_2$.
 Applying this reasoning recursively, we deduce  that  $W$ is connected. 
 We  emphasize that connected words
 remain connected after the use of   
 the simplification rules~(\ref{mi2}-\ref{mimj}).

\subsection{Calculation of $F_k$ for small $k$}

  We first remark  that \eqref{lntg} defines
   an infinity of operators $F_k$ but we have seen that 
there are only $L$ operators $H_k$ for a system of size $L$.  Therefore,  the
$F_k$'s  are not  all independent and  the knowledge of the $F_1, \dots, F_L$
is formally sufficient to generate all the  $F_k$.
Consequently, when we   consider $F_k$ in   the following formulae,
we   assume implicitly that the system is sufficiently  large to have
$k < L$. 
  The operator  $F_k$ is the $k$th order term in the expansion 
 of $\ln t_g(\lambda)$, given by equation~(\ref{fkhk}). After   
 using the  relations~(\ref{mi2})  and (\ref{mmm=0}),   $F_k$
 is found to be  a linear combination of words of length $j$, with $j \le k$.

For $k=1$,  $F_1$ is the Markov matrix $M$, 
\begin{equation}
  F_1 = H_1 = M = \sum_{i=1}^L M_i \, . 
  \label{f1}
\end{equation}
For $k=2$, we have 
\begin{equation}
  F_2 = 2 H_2 - H_1^2 = \sum_{i=1}^L ( [M_i, M_{i+1}] + M_i ) \ ,
  \label{f2}
\end{equation}
where we use the convention $M_{i+L} = M_i$ due to periodic boundary
conditions.  The operator  $F_2$ is indeed  connected~:
all non-connected terms in  $2 H_2 - H_1^2 $  of the type
 $M_i M_j$ with $|i-j| \ge 2$  cancel one another  and there remains only 
words of the type  $M_i M_{i+1}$ and $M_{i+1} M_i$,  involving the 
 adjacent  bonds  $(i,i+1)$ and $(i+1,i+2)$.

After an explicit  calculation,
   we find the following formulae for   $F_3$,  $F_4$
 and $F_5$~:
\begin{eqnarray}
  F_3 &=& 3 H_3 - 3 H_2 H_1 + H_1^3
\nonumber \\
      &=& \sum_{i=1}^L ( [[M_i, M_{i+1}], M_{i+2}] 
        + M_i M_{i+1} - 2 M_{i+1} M_i + M_i ) \ ;\label{f3}
\\
  F_4 &=& 4 H_4 - 4 H_3 H_1 - 2 H_2^2 + 4 H_2 H_1^2 - H_1^4
\nonumber \\
      &=& \sum_{i=1}^L \Big\{ [[[M_i, M_{i+1}], M_{i+2}], M_{i+3}] 
\nonumber \\ &&
          + M_i M_{i+1} M_{i+2} 
        - 2 ( M_{i+1} M_i M_{i+2} 
          + M_{i+2} M_i M_{i+1})
        + 3 M_{i+2} M_{i+1} M_i 
\nonumber \\ &&
        + M_i M_{i+1} - 3 M_{i+1} M_i + M_i \Big\} \ ;
\label{f4}
\\
  F_5 &=& 5 H_5 - 5 H_4 H_1 - 5 H_3 H_2 + 5 H_3 H_1^2 + 5 H_2^2 H_1
          - 5 H_2 H_1^3 + H_1^5 
\nonumber \\
      &=& \sum_{i=1}^L \Big\{ [[[[M_i, M_{i+1}], M_{i+2}], M_{i+3}], M_{i+4}]
\nonumber \\ &&
        + M_i M_{i+1} M_{i+2} M_{i+3} 
\nonumber \\ &&
      - 2 ( M_{i+1} M_i M_{i+2} M_{i+3}
        +  M_{i+2} M_i M_{i+1} M_{i+3}
        +  M_{i+3} M_i M_{i+1} M_{i+2} )
\nonumber \\ &&
      + 3 ( M_{i+2} M_{i+1} M_i M_{i+3}
      +  M_{i+3} M_{i+1} M_i M_{i+2}
      +  M_{i+3} M_{i+2} M_i M_{i+1} )
\nonumber \\ &&
      - 4 M_{i+3} M_{i+2} M_{i+1} M_i 
\nonumber \\ &&
        + M_i M_{i+1} M_{i+2} 
        - 3( M_{i+1} M_i M_{i+2} 
           +  M_{i+2} M_i M_{i+1} )
        + 6 M_{i+2} M_{i+1} M_i 
\nonumber \\ &&
        + M_i M_{i+1} - 4 M_{i+1} M_i + M_i \Big\} \ .
\label{f5}
\end{eqnarray}
 As expected, $F_k$ is made only of connected words. 
  We notice  the following remarkable
  property from   the expressions~(\ref{f3}-\ref{f5})
 ~: the  words of  length $j$  in  $F_k$
are always a permutation of $j$ consecutive matrices,
$M_i, M_{i+1}, \dots, M_{i+j-1}$, {\em without} repetition. 
  For example,  the expression~(\ref{f4})  of $F_4$ does not contain
  the word  $M_{i+1} M_i M_{i+2} M_{i+1}$. This property of $F_k$
 has been verified explicitly  for $k \le 10$. 

\subsection{A formula for the connected operators}

We have written a computer program that  gives the expressions
  of the $F_k$'s  for
small values of $k$ (up to $k =10$).
  This  leads  us to conjecture a  general
formula for $F_k$   valid  for arbitrary $k$. In  order to write
 this general formula  we  need to define some notations.

 \subsubsection{Simple words}

  A {\bf simple word
of length $j$} is defined as  a word $M_{\sigma(1)} M_{\sigma(2)} \dots
M_{\sigma(j)}$, where $\sigma$ is a permutation on the set $\{1,2,
\dots, j\}$. For example, there is a unique 
 simple word of length 1, noted $W_1 = M_1$ and two
simple words of length 2,  $W_2(1) = M_1 M_2$ and $W_2(0) = M_2
M_1$. For  $j \ge 2$,  the  commutation  rule~(\ref{mimj}) implies that 
  only the relative position of $M_i$  with 
  respect to $M_{i \pm 1}$ matters~:
 the number of  simple words of length $j$ is therefore much smaller
 than $j!$.
 In fact, any  simple word $W_j$ is uniquely   characterized by
  $(s_2, s_3, \dots, s_j)$
  where $s_i = 1$ if $M_i$ is written to  the right of $M_{i-1}$ in
$W_j$  and $s_i = 0$ otherwise.
Therefore, there are $2^{j-1}$ simple words of length $j$ and we note
them $W_j(s_2, s_3, \dots, s_j)$.  Simple words  obey the recursive
rule:
\begin{eqnarray}
   W_j(s_2, s_3, \dots, s_{j-1},1) &=& 
         W_{j-1}(s_2, s_3, \dots, s_{j-1}) \ M_j \ ,
\\
   W_j(s_2, s_3, \dots, s_{j-1},0)
    &=& M_j W_{j-1}(s_2, s_3, \dots, s_{j-1}) \, . 
\end{eqnarray}
 The set  of simple words of length $j$ will be called
   $\mathcal{W}_j$.

  For a   simple word $W_j$,  we define  $u(W_j)$ to  be the number of {\em
inversions} in  $W_j$, {\it i.e.}, the number of times
 that  $M_i$ is  on the left of $M_{i-1}$~: 
\begin{equation}
   u(W_j(s_2, s_3, \dots, s_j)) = \sum_{i=2}^{j} (1-s_i) \ .
\end{equation}
By definition, $0 \le u(W_j) \le j-1$.  
 For  example,  we have $W_5(1,0,1,0) = M_5 M_3 M_1
M_2 M_4$ and  $u(W_5(1,0,1,0))=2$.

 Using these definitions,  the 
  ``nested'' commutator  that appears in the 
  expressions~(\ref{f2}-\ref{f5}) 
 can be written   for general $k$ as~:
\begin{equation}
   [[ \dots [[M_1, M_2], M_3], \dots], M_k] = 
   \sum_{W \in \mathcal{W}_k} (-1)^{u(W)} W \, , 
\end{equation}
  where  $\sum_{W
\in \mathcal{W}_k}$ is equivalent to 
  writing  $\sum_{s_2=0}^1 \dots
\sum_{s_k=0}^1$.

\subsubsection{Conjectured general formula for $F_k$}

  We have calculated the exact  expressions of the connected
 operators up to $F_{10}$ and we  have 
  noticed  that  in  $F_k$  all  simple words  $W$ of length $j \le k$ appear
 with the sign $(-1)^{u(W)}$ and with a coefficient  given by 
  the binomial coefficient 
$\binomial{ k-j+u(W)}{k-j}$. Therefore, for $k<L$, we conjecture 
 the following general formula for $F_k$~:
\begin{equation}
  F_k = \mathcal{T} \sum_{j=1}^k \sum_{W \in \mathcal{W}_j}
            (-1)^{u(W)} \binomial{k-j+u(W)}{k-j} W \ ,
  \label{fk}
\end{equation}
 where $\mathcal{T}$ is the translation-symmetrizator 
 that  acts on any operator $A$ as follows~:
 \begin{equation}
   \mathcal{T} A  = \sum_{i=0}^{L-1}  t(0)^{i} \ A \  t(0)^{-i}\,.
 \end{equation}
  The presence of   $\mathcal{T}$   in equation~(\ref{fk})
  insures that  $F_k$ 
  is invariant by translation on the periodic system of
  size $L$.

   We also verified that  for $j+k \le 11$
  the conjecture (\ref{fk}) gives $[F_k,F_j] = 0$.  We emphasize  that
 because of the special expression~(\ref{hl}) of $H_L$
the expression (\ref{fk}) of $F_k$ is valid only for systems with length
$L > k$.

\subsection{Action of $F_k$ on a  configuration}

    In this section we  describe the action of $F_k$,
 as  given by the  formula~(\ref{fk}),  on an arbitrary  configuration
 $C = |\tau_1, \tau_2, \dots, \tau_L \rangle $. 
 We first define  an  operator   $A$, that we  shall call  
  the `Antisymmetrizator', by describing its action on  a configuration.
  The antisymmetrizator  $A$ acts  on a bond as follows~:
\begin{eqnarray}
   A  \ |01\rangle  &=&  |01\rangle -  |10\rangle  \, ,\\
  \hbox { and } 
  A  \ |\tau_i, \tau_{i+1} \rangle  &=&  |\tau_i, \tau_{i+1} \rangle \ ,
\,  \hbox{  for  } \, 
\tau_i \neq 0  \,  \hbox{ and } \tau_{i+1} \neq 1 \, . 
\end{eqnarray}
 More generally, 
  the action of $A$  is given by~: 
\begin{eqnarray}
 & A \ |1^{p_1}  0^{h_1+1}  1^{p_2+1}  0^{h_2+1}  1^{p_2+1}  \dots
           0^{h_{r-1}+1}  1^{p_r+1}  0^{h_r} \rangle = 
& \nonumber \\ & 
     |1^{p_1} 0^{h_1} \rangle \otimes 
      A \ |0 1\rangle
    \otimes |1^{p_2}  0^{h_2}\rangle \otimes  A \ |0 1\rangle
    \otimes \dots  \otimes 
      A \ |0 1\rangle
    \otimes |1^{p_r} 0^{h_r}\rangle \, , 
& \label{eq:defA}
\end{eqnarray}
 where  $ h_i, p_i \ge 0$. 

 Consider now a simple word $W_j(s_2, s_3, \dots, s_j)$  
 acting on a system of size $L > j$. This operator affects
 only the sites  $1,2\dots, j, j+1$, the sites $j+2, \ldots, L$
 being spectators.
   We show in Appendix~\ref{appendix:simple-word} that  
\begin{equation}
  W_j(s_2, s_3, \dots, s_j) \ 
  | \tau_1, \tau_2, \dots, \tau_L \rangle \ne  0
  \label{st}
\end{equation}
if and only if
\begin{equation}
  \tau_1 = 1, \ \tau_{j+1} = 0 \ \mbox{and} \ 
  \tau_2 =s_2, \ \tau_3 = s_3, \dots, \tau_j = s_j \ .
  \label{ts}
\end{equation}
  If this condition is satisfied, the action of the simple word is given by 
\begin{eqnarray}
  W_j(s_2, s_3, \dots, s_j)
  | 1, s_2, \dots, s_j, 0, \tau_{j+2}, \dots, \tau_L \rangle
  =  &&  \nonumber \\
  A  \, | 0, s_2, \dots, s_j, 1 \rangle \ \otimes
   \  |\tau_{j+2}, \dots, \tau_L \rangle   \, , && 
  \label{wc}
\end{eqnarray}
 where  $A$  is defined in equation~(\ref{eq:defA}). 
 Thus,  a word acts only on  specific configurations.
 From this remark, we can derive  a formula for the action of $F_k$ 
  on a  configuration $C$.  From  equation~(\ref{fk}),
   we first  observe that only one specific
  word  $W \in \mathcal{W}_j$ has a non zero action on
  a given  configuration   $C$~:
\begin{eqnarray}
  \lefteqn{
  \sum_{W \in \mathcal{W}_j}
            (-1)^{u(W)} \binomial{k-j+u(W)}{k-j} W
 | 1, \tau_2, \dots, \tau_j, 0, \tau_{j+2}, \dots, \tau_L \rangle
 = 
 } \nonumber \\ &&
(-1)^u \binomial{k-j+u}{k-j} 
  A | 0, \tau_2, \dots, \tau_j, 1 \rangle \ \otimes
   \  |\tau_{j+2}, \dots, \tau_L \rangle \, , 
\end{eqnarray}
where  $u = \sum_{i=2}^j (1-\tau_i)$ is the number of holes in 
$C$ between sites  1 et $j+1$. We emphasize that $u$ is 
 a function of $C$  only.  Now,  according to 
 equation~(\ref{fk}), we have to take a sum over $j$ and 
 apply the translation-symmetrizator  $\mathcal{T}$. This amounts  
  to  considering all possible jumps
 from  an occupied site  $i$   to
 an empty site   $m$ with  $ j = m -i \le k$. We thus obtain 
\begin{eqnarray}
  \lefteqn{
   F_k | \tau_1, \tau_2, \dots, \tau_L \rangle = 
 }  \nonumber  \\ &&
   \sum_{\stackrel{\scriptstyle i=1}{\tau_i=1}}^L
   \sum_{\stackrel{\scriptstyle m=i+1}{\tau_m=0}}^{i+k}
   (-1)^{u(i,m)} \binomial{k+i-m + u(i,m)}{k+i-m} \times  \nonumber
\\&&
   | \tau_1, \dots, \tau_{i-1} \rangle 
   \ \otimes \ 
   A | 0, \tau_{i+1}, \dots, \tau_{m-1}, 1 \rangle
   \ \otimes \ 
   | \tau_{m+1}, \dots, \tau_L \rangle \, , 
\end{eqnarray}
where  $u(i,m) = \sum_{r=i+1}^{m-1} (1-\tau_r)$ is  the number of holes in
   $C$ between  sites $i$ and $m$ (we recall that  
  sites are defined  modulo $L$).

 The action of  $F_k$  can be described as follows. 
  Each particle, starting  from an occupied site,   can 
 make   all possible jumps
 of length  $j \le k$   to  a vacant  site.  Each jump has a sign and
 a  weight~: 
 the  sign  is given by $(-1)^u$  where $u$ is the number of holes
 overtaken by the particle between its initial and its final position; 
 the weight is a binomial coefficient  that 
  depends only  on  $u$ and  $k-j$.

\section{Conclusion}

   The algebraic Bethe Ansatz technique allows to construct a family
 of operators that commute  with a given  integrable  hamiltonian.
  For the totally asymmetric  exclusion process, this procedure
  has enabled  us to define a family of generalized operators,
  local  and non-local, that commute with the Markov matrix.
 The properties of these operators 
  have been derived by using  the TASEP algebra~(\ref{mi2}-\ref{mimj})
 and their  actions   on the configuration space has
 been explicitly described.
   In particular, we  have found   a  combinatorial 
   formula for  the connected operators  valid at all orders.
   This formula has been verified  for systems of small size but 
   a mathematical proof remains to be established. 

   It would be of interest to extend
  our results to the  exclusion process with 
  forward and backward  hopping  rates. 
  Because   the symmetric exclusion
 process is equivalent to the Heisenberg spin chain, 
 the generalized hamiltonians would correspond to integrable models
   with long range interactions.  Explicit formulae for the connected
 conserved operators associated with the Heisenberg spin chain
are  known  only for the lowest orders (Fabricius et al., 1990);  
  no general expressions
 for these spin chain operators have  yet  been   found.
  We believe that the expression given in the present work, 
  equation~(\ref{fk}), that  is  valid at all orders,
  may shed some light on this issue. 
   
  Finally, we   hope that the  family of commuting  operators 
  studied in the present work 
 will help  to explain  the spectral degeneracies
 found in the Markov matrix and to unveil hidden algebraic symmetries
 of the exclusion process.

\subsection*{Acknowledgements}

We express our gratitude to T. Jolicoeur  and  to M. Bauer for
numerous helpful discussions and S. Mallick for a careful
 reading of the manuscript. 

\appendix

\section{Construction of the transfer matrix}
\label{appendix:aba}

In this appendix, we use the algebraic Bethe ansatz method to construct
the transfer matrix of the TASEP, a family $t(\lambda)$ of commuting
operators acting on the configuration space $\mathcal{H}_L$.

  \subsection*{Generalized  jump operators}

In section~\ref{section:mi}, we have defined $M_i$, the jump operator from
site $i$ to site $i+1$.  More generally, for two different sites $a$ and
$b$, we define $P_{a,b}$, the permutation operator between sites $a$ and
$b$  by
\begin{equation}
  P_{a,b} |\dots \tau_a, \dots \tau_b, \dots \rangle 
       =  |\dots \tau_b, \dots \tau_a, \dots \rangle \ ,
\end{equation}
and $M_{a,b}$, the jump operator from $a$ to $b$, by
\begin{equation}
  M_{a,b} = (P_{a,b} - 1 ) \ \sigma_a^{(1)} \sigma_b^{(0)}
  \label{mab}
\end{equation}
where $\sigma_i^{(\tau)} = |\tau_i \rangle \langle \tau_i |$  is  the
projector on the subspace of configurations with site $i$ in state
$\tau$. The operators  
 $M_{a,b}$ and $P_{a,b}$ act non trivially only on the subspace
$V_a \otimes V_b$  and   are  the identity operator  on all spaces
$V_i$ for $i$ different from  $a$ and  $b$.
 The  relations~(\ref{mi2}--\ref{mimj})  now become
\begin{eqnarray}
  M_{a,b}^2 &=& - M_{a,b} 
  \label{mab2}
\\
  M_{a,b} \ M_{b,c} \ M_{a,b} =
  M_{b,c} \ M_{a,b} \ M_{b,c} & = & 0 
\\
  \left[ M_{a,b}, M_{c,d} \right] &=& 0  
  \label{mabmcd}
\end{eqnarray}
where $a$, $b$, $c$ and $d$ are different sites.
  Equation~(\ref{mab})  allows  to define a totally asymmetric
 exclusion process  on an
arbitrary graph, with one jump matrix  $M_{a,b}$ for each directed edge
$(a,b)$ of the graph.  Consequently $M_i$ is just a simplified  notation for
$M_{i,i+1}$ when the graph is   a ring.

As the main problem is the non-commutativity of the neighboring jump
operators $M_i$ and $M_{i+1}$, the key step consists of finding
operators which verify a quasi-commutation rule, the Yang-Baxter
equation.  Such operators are given  by
\begin{equation}
   \mathcal{L}_{a,b}(\lambda) = P_{a,b} ( 1 + \lambda M_{a,b} ),
   \label{lab}
\end{equation}
where $a$ and $b$ are two given sites, and $\lambda$ is a number (the
spectral parameter).
The $\mathcal{L}_{a,b}$ satisfy 
  the Yang-Baxter equation (for a derivation see, e.g.,
 Golinelli and Mallick 2006b) ~: 
\begin{equation}
  \mathcal{L}_{a,b}(\nu) \mathcal{L}_{c,b}(\lambda) \mathcal{L}_{c,a}(\mu)
  =
  \mathcal{L}_{c,a}(\mu) \mathcal{L}_{c,b}(\lambda) \mathcal{L}_{a,b}(\nu) 
  \label{yb} \,\, \hbox{ if } \,\,\, \lambda = \mu + \nu - \mu \nu. 
\end{equation}

\subsection*{The monodromy matrix $\hat{T}(\lambda)$}

   To the $L$  physical  sites ($i=1,\dots, L$), we add an
 {\it auxiliary}  site with label 0. The extended 
  configurations are noted  as 
$ | \tau_0 \rangle \otimes | \tau_1, \dots, \tau_L \rangle$, with
$\tau_i \in \{0, 1\}$ for $i=0,\dots, L$,  and the extended 
  $2^{L+1}$ dimensional 
 state  space  is given by  $V_0 \otimes \mathcal{H}_L$. 
{\it In order to distinguish the spaces on which operators act, we note with
a ``hat'' $(\ \hat{} \ )$ the operators acting on the extended  space
$V_0 \otimes \mathcal{H}_L$, and without hat those  acting on the
physical space $\mathcal{H}_L$.} 

We define the {\em monodromy} matrix $\hat{T}(\lambda)$ by
\begin{equation}
  \hat{T}(\lambda) = \hat{\mathcal{L}}_{1,0}(\lambda) \ 
                     \hat{\mathcal{L}}_{2,0}(\lambda)
              \dots \ \hat{\mathcal{L}}_{L,0}(\lambda)
  \ . 
  \label{t}
\end{equation}
 The matrix  $\hat{T}(\lambda)$ acts on the extended space  
 $V_0 \otimes \mathcal{H}_L$. 
  We now consider two auxiliary sites  $0$ et $0'$ 
 and two monodromy matrices $T_0(\lambda)$ and $T_{0'}(\mu)$ acting
 on the  space $V_0 \otimes V_{0'} \otimes \mathcal{H}_L$. 
  Using  equation~(\ref{yb}) and the fact that 
  $[\mathcal{L}_{i,0}(\lambda), \mathcal{L}_{j,0'}(\mu)] = 0$
 for  $i \ne j$, 
 we deduce that $T_0(\lambda)$ and $T_{0'}(\mu)$
 also satisfy the  Yang-Baxter relation~:
\begin{equation}
  \mathcal{L}_{0',0}(\nu) \ T_0(\lambda) \ T_{0'}(\mu) 
  = 
  T_{0'}(\mu) \ T_0(\lambda) \ \mathcal{L}_{0',0}(\nu)
   \,\, \hbox{ if } \,\,\, \lambda = \mu + \nu - \mu \nu  \, .
 \label{ltt=ttl}
\end{equation}

 Using  definitions~(\ref{lab}, \ref{t})  we find  for $\lambda = 0$,
\begin{equation}
    \hat{T}(0)  =  \hat{P}_{1,0} \ \hat{P}_{2,0} \dots \ \hat{P}_{L,0}.
\end{equation}
 The explicit action of $\hat{T}(0)$ on an extended configuration is then  
\begin{equation}
  \hat{T}(0) 
 ( | \tau_0 \rangle \otimes | \tau_1, \tau_2 \dots, \tau_L \rangle) 
  = | \tau_1 \rangle \otimes | \tau_2 \dots, \tau_L, \tau_0 \rangle. 
  \label{t0}
\end{equation}
It turns out that $\hat{T}(0)$ is the {\bf translation}
  operator which causes a
left circular shift of the sites, including the auxiliary site 0.

In \eqref{t} for a generic $\lambda$, we can ``push''  the
permutation operators $\hat{P}_{i,0}$ to the left  using the relation
$\hat{M}_{i-1,0} \ \hat{P}_{i,0} = \hat{P}_{i,0} \ \hat{M}_{i-1,i}$, and obtain
\begin{equation}
   \hat{T}(\lambda) =  \hat{T}(0) ( 1 + \lambda \hat{M}_{1,2} ) 
                                  ( 1 + \lambda \hat{M}_{2,3} )
                            \dots ( 1 + \lambda \hat{M}_{L-1,L} )  
                                  ( 1 + \lambda \hat{M}_{L,0} ) \, .
\end{equation}
The operator  $\hat{T}(\lambda)$ is a  polynomial of degree $L$,
\begin{equation}
   \hat{T}(\lambda) = 
     \hat{T}(0) \ \left( 1 + \sum_{k=1}^L \lambda^k \hat{T}_k \right)
   \label{t-t0-tk}
\end{equation}
where the $\hat{T}_k$'s, that  act  on  $V_0 \otimes
\mathcal{H}_L$, are  given by
\begin{equation}
   \hat{T}_k = \sum_{1 \le i_1 < i_2 < \dots < i_k \le L} 
          \hat{M}_{i_1, i_1+1} \ \hat{M}_{i_2, i_2+1} \dots  
          \hat{M}_{i_k, i_k+1}
  \label{tk}
\end{equation}
for $ 1 \le k \le L$, with the convention $\hat{M}_{L,L+1} \equiv
\hat{M}_{L,0}$ when $i_k = L$.
 Hence, the operator 
 $\hat{T}_k$   represents the simultaneous jumps of  $k$ different particles
 initially located on  the physical sites.  In particular,
$\hat{T}_1$ is the Markov matrix of the TASEP on the open segment $(1,
2, \dots, L, 0)$.

\subsection*{The trace over the auxiliary space}

As the  operators defined  above  act on the extended space $V_0 \otimes
\mathcal{H}_L$, we will use the partial trace
$\tr_0$ over the auxiliary  space
 $V_0$  to obtain operators
acting only on the physical space $\mathcal{H}_L$.
 Any operator $\hat{A}$ acting on $V_0 \otimes \mathcal{H}_L$ can
be  uniquely written  as 
\begin{equation}
  \hat{A} = \sum_{\tau_0, \tau'_0 = 0}^1 
     \left( |\tau_0\rangle \langle \tau'_0 | \right)
     \otimes A(\tau_0, \tau_0')
\end{equation}
where $A(\tau_0, \tau_0')$ is an operator acting on $\mathcal{H}_L$.
The  partial trace is defined as
\begin{equation}
   \tr_0 \ \hat{A} = \sum_{\tau_0=0}^1 A(\tau_0, \tau_0) = A(0,0) +
   A(1,1) \, , 
\end{equation}
 and the  action  of $ \tr_0 \ \hat{A}$  is  given by 
\begin{eqnarray}
    \tr_0 \ \hat{A} | C \rangle =
   \sum_{\tau_0=0}^1
     \langle \tau_0 |  \hat{A}
   \left( |\tau_0 \rangle \otimes | C \rangle \right)
 \label{deftrace1} \, ,  \\
    \langle C'| \tr_0 \ \hat{A}  =
   \sum_{\tau_0=0}^1
   \left( \langle \tau_0 | \otimes \langle C'| \right) \hat{A} |\tau_0 \rangle 
 \label{deftrace2} \, .
\end{eqnarray}

Another property of the trace that we shall need is the following.
 Consider an operator $\hat{X}$ that acts only on $\mathcal{H}_L$;
 this operator can thus be written as  $\hat{X} = 1 \otimes X$.
 Then for any  $\hat{A}$ acting on $V_0 \otimes \mathcal{H}_L$
 we have~:
 \begin{eqnarray}
   \tr_0 ( \hat{A} \hat{X}) &=&   \tr_0 ( \hat{A})\  X  \, , \label{trdte} \\
     \tr_0 ( \hat{X} \hat{A}) &=&  X  \ \tr_0 ( \hat{A})  \, . \label{trgauche}
 \end{eqnarray}

\subsection*{The transfer matrix $t(\lambda)$}

The {\em transfer matrix} $t(\lambda)$, which 
 acts   on the physical configuration space $\mathcal{H}_L$  is defined by
\begin{equation}
     t(\lambda) = \tr_0 \ \hat{T}(\lambda) \, . 
     \label{t-tr}
\end{equation}
  The operators $\mathcal{L}_{i}(\lambda)$ and  the
monodromy matrix $\hat{T}(\lambda)$  conserve the number of particles in 
  the extended  space (physical space  plus the auxiliary site). 
  As the auxiliary trace $\tr_0$ operation keeps
constant the number $\tau_0$ of particles on the auxiliary site,  it
keeps   the number of particles in the physical space constant too.
Hence by construction, the transfer matrix $t(\lambda)$ conserves the
number of particles.

  We now multiply the  relation~(\ref{ltt=ttl}) by
$\mathcal{L}_{0',0}^{-1}(\nu)$ on the left
 and take its  trace   $\tr_{0,0'}$ over the two auxiliary 
 sites  0 and  0'.  Because  $\mathcal{L}_{0',0}^{-1}(\nu)$
 acts only on   0 and  0', we can use  that  
  $\tr_{0,0'}$  is   cyclic  with respect to  $\mathcal{L}_{0',0}^{-1}(\nu)$
 and thus 
\begin{equation}
  \tr_{0,0'}[ T_0(\lambda) \ T_{0'}(\mu) ] =
  \tr_{0,0'}[ T_{0'}(\mu) \ T_0(\lambda) ] \, .
\end{equation}
 Using~(\ref{t-tr}) and the relation
  $\tr_{0,0'} = \tr_0 \ \tr_{0'}$,  we obtain 
\begin{equation}
  t(\lambda) \ t(\mu) = t(\lambda) \ t(\mu) \, .
  \label{tt=tt}
\end{equation}
 The Yang-Baxter equation~(\ref{yb}) thus implies the commutativity
 of the transfer matrices.

\section{Calculation of the  hamiltonians $H_k$}
\label{app:calculHk}

  We   derive  here the expression~(\ref{hk})  for  $H_k$.
Following Eqs.~(\ref{thk}, \ref{t-t0-tk}, \ref{t-tr}), we obtain,
 for $ 1 \le k \le L$, 
\begin{equation}
  t(0) =  \tr_0\  [\hat{T}(0)]   \, 
  \hbox{ and } \,  H_k = t(0)^{-1} \ \tr_0\  [\hat{T}(0) \hat{T}_k] \, .
  \label{hk-tr}
\end{equation}
 We shall now perform  the trace $\tr_0$  over the auxiliary space.

 We first  calculate  $t(0)$~: for 
 a given configuration $ | \tau_1, \tau_2, \dots, \tau_L \rangle $ of
the physical sites, we obtain using Eqs~(\ref{deftrace1}) and
(\ref{t-tr})
\begin{equation}
  t(0) | \tau_1, \tau_2, \dots, \tau_L \rangle = 
  \sum_{\tau_0 = 0}^1 \langle \tau_0 | \hat{T}(0) 
     ( \, | \tau_0 \rangle \otimes | \tau_1, \tau_2, \dots, \tau_L \rangle )
 \, .   \label{t0t0t0}
\end{equation}
  As $\hat{T}(0)$ is the
translation operator on the extended  space, we obtain 
\begin{equation}
    t(0) | \tau_1, \tau_2, \dots, \tau_L \rangle = 
    \sum_{\tau_0 = 0}^1 \langle \tau_0 |
    ( \, | \tau_1 \rangle \otimes | \tau_2, \dots, \tau_L, \tau_0
    \rangle ).
\end{equation}
In the auxiliary space $V_0$, we have $\langle \tau_0 | \tau_1
\rangle = \delta_{\tau_0,\tau_1}$ and  then
\begin{equation}
   t(0) | \tau_1, \tau_2, \dots, \tau_L \rangle =
   | \tau_2, \dots, \tau_L, \tau_1 \rangle \ .
   \label{pt0app}
\end{equation}
Therefore   $t(0)$ is the {\em
translation operator} on the configuration space.

  We now evaluate $H_k$ for $1\le k \le L-1$.
 According to Eqs.~(\ref{tk}, \ref{hk-tr}), any
 term $W$  that appears in $\hat{T_k}$  
  is made of $k$ jump operators with $k < L$.
 Thus,   such a term $W$
  can  always be  written as   $ \hat{D}\hat{F}$  with 
\begin{eqnarray}
 \hat{D}  &=&  \hat{M}_{i_1, i_1+1} \ \hat{M}_{i_2, i_2+1} \dots 
 \hat{M}_{i_r, i_r+1}  \,\,\, 
  \hbox{with}  \,\,\,   1 \le i_1 < i_2 < \ldots < i_r \le  u-1 \, ,
 \nonumber \\
    \hat{F} &=&  \hat{M}_{i_{r+1}, i_{r+1}+1}  \dots  
          \hat{M}_{i_k, i_k+1}
  \,\,\, \hbox{with}  \,\,\,  u+1 \le i_{r+1} < \ldots < i_k \le L  \, . 
  \,\,\, \label{def:DF}
\end{eqnarray}
 The index $u$ is such that the matrix $\hat{M}_{u, u+1}$
  does not appear in $W$.
 Therefore, all the traces 
 that we have to  calculate  are of the type 
$\tr_0(\hat{T}(0)\hat{D}\hat{F})$ with $[\hat{D},\hat{F}]=0$.
 Besides we notice that $\hat{D}$ acts only on $\mathcal{H}_L$. Therefore,
 we have, using equation~(\ref{trdte})~:
\begin{equation}
 \tr_0(\hat{T}(0)\hat{D}\hat{F}) =  \tr_0(\hat{T}(0)\hat{F}\hat{D})
                                =  \tr_0(\hat{T}(0)\hat{F}) D \, , 
  \label{traTDF1}
\end{equation}
 with $D = M_{i_1}M_{i_2}\ldots M_{i_r}$.
The operator $\hat{F}$ can not be extracted from the trace because
 it  acts on the auxiliary site 0 if $i_k =L$ in   equation~(\ref{def:DF}).
 However, recalling that  $\hat{T}(0)$ is the
 translation operator on the total space  $V_0 \otimes \mathcal{H}_L$,
 we can write 
\begin{equation}
\hat{T}(0)\hat{F} =  \hat{F}'\ \hat{T}(0) \ \hbox{ with } \
 \hat{F}'  =  \hat{M}_{i_{r+1}-1, i_{r+1}}  \dots  
          \hat{M}_{i_k -1, i_k} \, .
\end{equation}
 The fact that  $u \ge 1$ and $i_k \le L$ ensures  that $\hat{F}'$
acts only on  $\mathcal{H}_L$ and as such can be written as
 \begin{equation}
 \hat{F}' = 1 \otimes F' \ \hbox{with} \ 
 F' = M_{i_{r+1}-1} \ldots M_{i_k -1} \, . \label{def:F'}
 \end{equation}
 We  now use  the property~(\ref{trgauche})
 and write  equation~(\ref{traTDF1}) as follows
\begin{equation}
 \tr_0(\hat{T}(0)\hat{D}\hat{F}) = 
   \tr_0( \hat{F}'  \hat{T}(0)) D = F' 
   \tr_0(\hat{T}(0))  D = F' t(0) D\, .
\end{equation}
 Using the fact that $t(0)$ is the translation operator on the configuration 
space, we write
 \begin{equation}
 F' t(0) =  t(0) F \ \hbox{with} \  F = M_{i_{r+1}}\ldots M_{i_k} \, , 
\end{equation}
 and   conclude that
 \begin{equation}
 \tr_0(\hat{T}(0)\hat{D}\hat{F}) =  t(0)  F   D =  t(0) \rop{DF} \,, 
\end{equation}
 where $\rop{}$ is defined in section  \ref{section:op}.
 This proves the general formula~(\ref{hk}).

To be complete we need to  calculate the
  operator of  the highest degree $H_L$.  The operator
\begin{equation}
  \hat{T}_L = M_{1,2} \dots M_{L-1,L} \ M_{L,0}
\end{equation}
involves jumps from {\em all} physical sites~: it can not
 be split as described in equation~(\ref{def:DF}). 
 We have  $\hat{T}_L | C \rangle =
0$ for all configurations $C$, unless $ C = |0\rangle \otimes | 1,
1,\dots 1\rangle$.  After a short calculation, \eqref{hk-tr} leads to 
that
\begin{equation}
   H_L = | 1, 1, \dots, 1 \rangle   \langle 1, 1, \dots, 1| \ ,
\end{equation}
which  is  the projector on the ``full'' configuration (all sites are
occupied) in agreement with equations~(\ref{m1ml}) and (\ref{hl}).

\section{Derivation of Eq. (\ref{st}-\ref{wc})}
\label{appendix:simple-word}

 In this appendix, we prove equations~(\ref{st}-\ref{wc}) by induction
 on the size $j$ of the simple word $W = W_j(s_2, s_3, \dots, s_j)$.
 We shall simplify the notations by writing  the action of $W$ on 
  the sites  $1,2\dots, j, j+1$ (the sites $j+2, \ldots, L$
 being spectators).

  For $j=1$, the only word is  $W_1 = M_1$
 and  equations~(\ref{st}-\ref{wc})  are satisfied~:
\begin{eqnarray}
  M_1  |\tau_1,\tau_2\rangle \neq 0  \, \hbox { iff }  \, \tau_1 = 1, \, 
\tau_2 = 0 \,  \, \\  \hbox { and  }  \, 
  M_1 \ |1,0\rangle = |0,1\rangle - |1,0\rangle = A \ |0,1\rangle.
\end{eqnarray}

  For  $j \ge 2$, we shall calculate the action of the word  $W$  
  on the configuration 
   $C = |\tau_1,  \tau_2  \dots  \tau_{j+1}\rangle$.
  We must distinguish two cases  $s_j = 1$ or  0.

\subsubsection*{Case $s_j=1$}
   
  We can  write 
  $W = W' \ M_j$  where $W' = W_{j-1}(s_2, \dots, s_{j-1})$
  is a simple word of length $j-1$ and we have
\begin{equation}
  W \ |\tau_1  \dots \tau_{j-1}, \tau_j,  \tau_{j+1} \rangle
 =  W' \ M_j  \ |\tau_1  \dots  \tau_{j-1}, \tau_j,  \tau_{j+1} \rangle
\end{equation} 
 This action vanishes unless  $\tau_j = 1 = s_j$  and 
  $\tau_{j+1} = 0$. In that case we have 
\begin{equation}
  W \ |\tau_1  \dots  \tau_{j-1}, 1, 0  \rangle  
  = W' \ |\tau_1  \dots  \tau_{j-1},  0,  1 \rangle
  - W' \ |\tau_1  \dots  \tau_{j-1},  1,  0 \rangle \ .
\end{equation}
 We can now use the induction hypothesis~: the  second  term on the r.h.s.
 always vanishes (because $\tau_j = 1$); the first term on the r.h.s.
 does not vanish if $\tau_1 = 1$ and  $\tau_2 = s_2$, \dots, $\tau_{j-1}
= s_{j-1}$. Therefore, the action of $W$ on $C$ does not vanish
 if and only if  $C = |1, s_2 \dots s_{j-1}, 1, 0 \rangle$ 
 and   is given by
\begin{equation}
    W  \ |1, s_2 \dots s_{j-1}, 1, 0 \rangle
 =  W'  \ |1, s_2 \dots s_{j-1}, 0,  1 \rangle
  = A   \ |0,  s_2 \dots  s_{j-1},  1,  1\rangle   \, , 
\end{equation}
 where we have used the induction hypothesis to evaluate the
 action of  $W'$ (we recall  the  site number $j+1$ is spectator
 for  $W'$). 
 Equations~(\ref{st}-\ref{wc}) are thus  proved for  the case $s_j=1$.

\subsubsection*{Case $s_j=0$}

   We now have  $W = M_j \ W'$  where
  $W'$ is defined as above.
 Therefore
\begin{equation}
  W \ |\tau_1  \dots \tau_{j-1}, \tau_j,  \tau_{j+1} \rangle
 = M_j \,  W' \ |\tau_1  \dots  \tau_{j-1}, \tau_j,  \tau_{j+1} \rangle \, .
\end{equation} 
  The  induction  hypothesis implies that the action of 
 $W'$ does not vanish if and   only  if 
  $\tau_1 = 1$, $\tau_2 = s_2$, \dots, $\tau_{j-1}
  = s_{j-1}$, $\tau_j = 0 = s_j$,  the site
 $(j+1)$ being spectator for $W'$. Besides, the action of  
   $M_j$   on the bond $(j,j+1)$ is non-trivial 
 only   if $\tau_{j+1} = 0$. Therefore, we  have
 $C = |1,  s_2 \dots  s_{j-1},  0, 0 \rangle$ and 
\begin{equation} 
   W  \ |1, s_2 \dots s_{j-1}, 0,  0 \rangle 
 = M_j \ \left( \ A \ | 0,  s_2  \dots  s_{j-1}, 1 \rangle \otimes
 | 0 \rangle \ \right) \, .
 \label{cas2}
\end{equation}
 The action of $A$ on the r.h.s. of this equation is given by
\begin{eqnarray} 
A \ | 0,  s_2  \dots  s_{j-2},  1,  1\rangle
 &=& A \ | 0,  s_2  \dots  s_{j-2},  1\rangle    \otimes |1\rangle
 \, \hbox{ if }  s_{j-1} = 1 \, , \\
A \ | 0,  s_2  \dots  s_{j-2},  0,  1\rangle
  &=& A \ | 0,  s_2  \dots  s_{j-2}\rangle  \otimes A |0,  1\rangle
 \, \hbox{ if }  s_{j-1} = 0 \, .
 \end{eqnarray} 
 Thus, we obtain, if  $s_{j-1} = 1$,
\begin{eqnarray}
   W  \ |1, s_2 \dots s_{j-2}, 1,  0,  0 \rangle  &=& 
 A \ | 0,  s_2  \dots  s_{j-2},  1\rangle \otimes M_j |1,  0\rangle
 \nonumber  \\
     &=& A \ | 0,  s_2  \dots  s_{j-2},  1\rangle \otimes A |0,  1\rangle
    \nonumber \\  &=& A \ | 0,  s_2  \dots  s_{j-2},  1,  0 , 1\rangle \, ,
\end{eqnarray}
 and if  $s_{j-1} = 0,$
\begin{eqnarray}
  W  \ |1, s_2 \dots s_{j-2}, 0,  0,  0 \rangle  &=& 
    A \ | 0,  s_2  \dots  s_{j-2} \rangle  \otimes 
 M_j (A |0, 1\rangle \otimes|0\rangle)  \nonumber   \\
    &=& A \ | 0,  s_2  \dots  s_{j-2} \rangle  \otimes 
  |0\rangle \otimes A |0, 1\rangle   \nonumber   \\
    &=& A \ | 0,  s_2  \dots  s_{j-2},  0,  0,  1\rangle \, ,
\end{eqnarray}
 which completes  the proof of  \eqref{wc}.

\section*{References}

\begin{itemize}

\item
Arnaudon D., Cramp\'e N., Doikou A., Frappat L. and Ragoucy E., 2005,
{\em Analytical Bethe Ansatz for closed and open $gl(N)$-spin chains in
any representation}, 
J. Stat. Mech. {\bf 2005} P02007,
arXiv:math-ph/0411021.

\item
  Brankov J. G., Priezzhev V. B. and Shelest R. V., 2004,
  {\em Generalized determinant solution of the discrete-time totally
  asymmetric exclusion process and zero-range process}, 
  Phys. Rev. E {\bf 69} 066136.

\item de Gier~J.  and Essler~F.~H.~L., 2005,  
{\em Bethe Ansatz solution of the asymmetric exclusion process
 with open boundaries}, Phys. Rev. Lett.  {\bf  95}  240601.

\item de Gier~J.  and Essler~F.~H.~L., 2006, 
 {\em Exact spectral gaps of the asymmetric exclusion process 
 with open boundaries}, 
 J. Stat. Mech. {\bf 2006} P12011,
 arXiv:cond-mat/0609645. 

\item Derrida B., 1998, {\em An exactly soluble non-equilibrium
    system: the asymmetric simple exclusion process}, Phys. Rep.  {\bf
    301}  65.

\item Derrida B.  and Lebowitz J.~L., 1998, {\em Exact large deviation
    function in the asymmetric exclusion process}, Phys. Rev. Lett.
  {\bf 80}  209.

\item Dhar D.  1987, {\em An exactly solved model for interfacial
    growth}, Phase Transitions {\bf 9}  51.

\item 
Essler~F.~H.~L.  and  Rittenberg~V.,  1996, 
 {\em  Representations of the quadratic algebra and partially
 asymmetric diffusion with open boundaries}, 
 J. Phys. A: Math. Gen. {\bf 29}  3375.

\item  Fabricius~K., M\"utter~K.-H. and Grosse~H., 1990,
 {\em  Hidden symmetries in the one-dimensional antiferromagnetic
 Heisenberg model}, Phys. Rev. B {\bf 42}  4656.

\item 
Golinelli O. and Mallick K., 2004,
{\em Hidden symmetries in the asymmetric exclusion process},
J. Stat. Mech. {\bf 2004}  P12001,  arXiv:cond-mat/0412353.

\item 
Golinelli O. and Mallick K., 2005,
{\em Spectral degeneracies in the totally asymmetric exclusion process},
J. Stat. Phys {\bf 120} 779.  

\item
Golinelli O. and Mallick K.,  2006a, 
{\em Derivation  of a matrix product representation
  for the asymmetric exclusion process from    algebraic Bethe Ansatz}, 
 J. Phys. A: Math. Gen. {\bf 39} 10647. 

\item
Golinelli O. and Mallick K.,  2006b, 
{\em The asymmetric simple exclusion process~:
 an integrable model for  non-equilibrium statistical mechanics},
J. Phys. A: Math. Gen. {\bf 39} 12679.

\item Gwa L.-H. and  Spohn H., 1992, {\em Bethe solution for the
    dynamical-scaling exponent of the noisy Burgers equation}, Phys.
  Rev. A {\bf 46}  844.

\item Kim D., 1995, {\em Bethe Ansatz solution for crossover scaling
    functions of the asymmetric XXZ chain and the
    Kardar-Parisi-Zhang-type growth model}, Phys. Rev. E {\bf 52} 
  3512.

\item Korepin V.E., Bogoliubov N.M. and Izergin A.G., 1993,
{\em Quantum inverse scattering method and correlation functions},
University Press, Cambridge.

\item L\"uscher~M., 1976,  {\em Dynamical charges in the quantized
 renormalized massive Thirring model},  Nucl. Phys. B {\bf 117}  475.

\item  Nepomechie R. I., 1999,  
{\em A spin chain primer},
Int. J. Mod. Phys. B  {\bf 13}  2973,
arXiv:hep-th/9810032.

\item Rajesh~R. and  Dhar~D., 1998,   {\em An exactly solvable anisotropic
  directed percolation model in three dimensions},  
    Phys. Rev. Lett. {\bf 81}  1646.

\item
  Rajewsky N., Schadschneider A. and Schreckenberg M., 1996,
  {\em The asymmetric exclusion model with sequential update},
  J. Phys. A: Math. Gen. {\bf 29} L305.

\item
  R\'akos A. and Sch\"utz G. M., 2005, 
  {\em Current distribution and random matrix ensembles for an integrable
  asymmetric fragmentation process},
  J. Stat. Phys. {\bf 118} 511.

\item   Sch\"utz~G.~M., 1993,
 {\em Generalized  Bethe Ansatz solution of a one-dimen\-sional
  asymmetric exclusion  process on a ring with blockage},
 J. Stat. Phys. {\bf 71}   471.

\item   Sch\"utz~G.~M., 2001,
{\em Exactly solvable models for many-body systems far from equilibrium}
 in {\em Phase Transitions and Critical Phenomena vol. 19}, C. Domb and
 J.~L.~Lebowitz Ed., Academic Press, San Diego.

\item  Spohn~H., 1991,
{\em Large scale dynamics of interacting particles},
 Springer, New-York.

\end{itemize}

\end{document}